\ifx\documentclass\undeftex
\documentstyle[11pt]{article}
\else
\documentclass[11pt]{article}
\usepackage{latexsym}
\fi

\newcommand{\langfont}{\sc}       
\newcommand{\scriptlangfont}{\rm} 

\newcommand{\shortsatord}[1]%
{\mbox{\langfont ShortSat$_{\scriptlangfont ord}(#1)$}}

\begin{document}

\sloppy

\title{
Writing and Editing Complexity Theory: Tales and 
Tools\protect\footnote{\copyright\ Lane 
A. Hemaspaandra and Alan L. Selman, 1998.}}

\author{
Lane A. Hemaspaandra$\,$\footnote{Supported in part by 
grants NSF-CCR-9322513 and 
NSF-INT-9513368/DAAD-315-PRO-fo-ab.  Done in part while 
visiting 
Friedrich-Schiller-Universit\"at Jena.
Email: {\tt lane@cs.rochester.edu}.
Department of Computer Science, University of Rochester,
Rochester, NY 14627, USA.}
\and
Alan L. Selman$\,$\footnote{Email: {\tt selman@cs.buffalo.edu}.
Department of Computer Science,
University at Buffalo,
Buffalo, NY 14260,~USA.}}

\date{November 1, 1998}

\maketitle

\begin{abstract}
\noindent
Each researcher should have a full shelf---physical or virtual---of
books on writing and editing prose.  Though we make no claim to any
special degree of expertise, we recently edited a book of complexity
theory surveys~\cite{hem-sel:b:ctrII}, 
and in doing so we were brought into particularly
close contact with the subject of this article, 
and with a number of the excellent resources available to
writers and editors.  In this article, we list some of these
resources, and we also relate some of the adventures we had 
as our book moved from concept to reality.
\end{abstract}

\section{Introduction}
Currently, the lingua
franca of computer science is English.  This is
a somewhat perverse situation since English is a difficult language to
use skillfully.  Clear 
communication is so
central to science that whatever effort one puts into careful use of
the language will be amply rewarded.

Many scientists already have a collection of language
reference books (or a stuffed browser bookmarks folder).
Nonetheless, we 
hope that this article will be useful and serve as a consciousness-raising
reminder of the importance of writing well.    The
first part of the article tells what we learned as the editors of {\em
Complexity Theory Retrospective~II}.~~The second part of this article
provides an annotated list of some of the books that we refer to often
when writing or editing technical prose.

\section{Tales}

During the spring of 1995, some months before the tenth meeting of 
the annual IEEE~Structure in Complexity Theory Conference 
(now the IEEE~Conference on Computational Complexity), 
we  agreed that this 
would be an appropriate occasion for putting together a book much like  the 
earlier 
{\em Complexity Theory Retrospective}~\cite{sel:b:retrospective} 
that Alan edited 
and that Springer-Verlag published in 1990.
We contacted the computer
science editor at Springer-Verlag.  He liked the idea and invited us to send
him a formal proposal.  We wrote a proposal that expressed our desire 
to provide a forum for expository articles that would 
present the most exciting new subareas and the most important advances 
of the last half-decade, and that proposed authors who we 
believed would be up to the task.  
Springer-Verlag's editor liked this proposal and so he invited us to prepare
a hardcover book of about three hundred 
pages.  This was a conscious decision on 
his part.  Hardcover publication
conveys a sense of permanence and worthiness that
contributes to a person's desire to make a purchase.

It was generally understood that we were  going to engage in this project
as a labor of love.  This is not the type of book that make publishers 
wealthy, so we would be involved in all aspects of the process. 
Such aspects 
would include creating the index, editing the individual articles, and
processing the text files into final book format.  

{\em Complexity Theory Retrospective II}~\cite{hem-sel:b:ctrII} 
appeared in print in 1997.  This was about two 
years after we began, and was in time to display  
a copy  at the Philadelphia FCRC\@.  
We encountered several problems during these two years, and we now
describe some things we learned.

A professional at Springer-Verlag built the index for the earlier book,
{\em Complexity Theory Retrospective}.
She created separate author and subject indexes, so we wanted separate
indexes for authors and subjects in the new book also.  Springer's 
\LaTeX~guru
said that he would  write an addition to their style file to 
accommodate this wish.
In turn, we asked authors to attach \verb|\index| commands to all of their
subject terms and to label all of the names of authors that
they cite with the new command \verb|\aindex|. 
The guru was  eventually to create \verb|\aindex| to write 
all author index information into a separate file.  
Indexing is tricky stuff and several decisions need to be made.
Should it be flat or nested? If you were looking up ``garlic bread''
in a cookbook, would you search under ``garlic'' or under ``bread''?
We studied the sixty-page discussion on indexing in 
{\em The Chicago Manual of Style}~\cite{chi:b:manual-style}, 
and we settled on an essentially mixed style. 
Along the way, we learned that \LaTeX~and
the fonts at Spring-Verlag are not exactly what we have in
the Unix world of academia.  Also, Springer's \LaTeX~style file does some
funny unexpected things now and then.  For this reason, we
abandoned the desire for separate author and subject indexes.
As always, the more bells and whistles, the greater the likelihood of
something breaking down.  We made this decision rather late in the process, 
and would recommend leaving indexing to 
the professionals.  

After authors wrote their papers, we circulated each 
paper to two other contributors, 
who prepared 
``referee reports.''  All authors graciously revised their papers
in accordance with the comments that they received.  Then  
we edited the papers, a time-consuming process, after which we forwarded
each paper back to the authors once more for their final proofreading.  Later
in this article, we will describe some of the most common errors 
that we found and corrected.

We compiled the 
various author's files into a draft of the book and then  Springer-Verlag 
assigned
a production editor who would now be responsible for
overseeing the final stages of publishing.
We sent our book, as a collection of
\LaTeX~files, to the production editor.  He sent 
a copy of the book, in paper form, to Springer-Verlag's 
professional copy editor.
The copy editor marked the pages and then the production editor marked
the pages, using a different color pen than the copy editor 
and sometimes telling us to ignore one of the copy editor's 
comments.  Springer-Verlag did not write onto any of the files, 
as that was part of our obligation.  Thus, being very careful
about the logistics, the production editor sent to Lane marked copy of the
chapters for which Lane was handling the editing, 
sent to Alan marked copy of the chapters for which Alan was 
handling the editing (so that by the color 
coding, we could tell who wrote which comments), and sent to each of us
photocopies of each other's marked 
chapters.  The production editor was particularly
interested in maintaining a certain degree of consistency, difficult 
with a book that has twelve authors,  and with
correctness and completeness of the bibliographies.  The latter is 
particularly worth remembering and is important.  
``FOCS 94'' means nothing to
a librarian.  Even though we carefully read and edited each 
chapter, the copy editor's notes were still copious.

The general pattern was that the copy editor energetically,
if somewhat mechanically, enforced all standard
rules.  For example, the copy editor rewrote 
all split infinitives---even those few that were appropriate 
and needed to convey their sentences' meanings.  Some of the 
suggested changes were surprises to us.  For example, the 
copy editor suggested that the ``Karp-Lipton Theorem'' be 
changed to the ``Karp--Lipton Theorem'' unless Karp had married
Lipton and solely authored the theorem.   (We did not follow that 
particular suggestion.)

After we completed the editorial revisions that the copy and production 
editors asked for, we once again sent a copy of all
of our files to them.  Their \LaTeX~guru once again processed the book using
their commercial \LaTeX,~fonts, and printers.  Now the book was very close
to being complete.  They sent the result to us for final proofreading, 
asking us to correct, among other things, various overfill boxes.
This turned out to be  impossible, because, since
we did not have access to their \LaTeX~and their fonts, we could
not duplicate their overfill boxes.  It would be preferable for the publisher
to  write onto the files as needed at this stage, 
and then send copy to the editors for
final proofreading.  Those of you who are considering such a 
project should consider including this in your agreement with your publisher.

Let's return to the question of consistency:  Editors need to inform authors
of the theorem style they should use.  Authors should get into the habit of
never hardcoding cross-references of any kind;  one should always use 
\LaTeX's 
\verb|\label| and \verb|\ref| commands.  Similarly, never hardcode
citations to the bibliography.  
Always place reference sources into a bib file, refer to them
using the \verb|\cite| command, and use Bib\TeX~to create the bibliography.
Authors need to take the time to create complete bibliography records
that include fully spelled-out names of journals and proceedings.  
This takes time, but scholarship, if not editors, should demand nothing less.

\section{Tools}\label{s:tools}

Below is an annotated list of the books that we often refer 
to when writing and editing, sorted by 
type.  

\subsection{Dictionaries}

There are two major, (relatively) up-to-date, unabridged dictionaries
of American English: 
{\em Webster's Third New International 
Dictionary}~\cite{web:b:new-international-third-edition}
and
{\em The {R}andom {H}ouse Unabridged
Dictionary}~\cite{fle:b:random-dictionary}.
{\em The American Heritage Dictionary of the English
Language}~\cite{ah:b:dictionary}, though not unabridged, has its
fans.
However, if you really want to make language lovers fall to their
knees as they enter your office, find the {\em second \/} edition of
{\em Webster's New International
Dictionary}~\cite{web:b:new-international-second-edition}.  
With its rich selection of usage examples, it puts even the
current (third) edition to shame.  Unfortunately, the second edition
has long been out of print and is essentially impossible to find.
One's only hope is a used book store or a garage sale.

If what you want to bring to its knees is your budget, there is
always the 20-volume {\em Oxford English Dictionary}~\cite{oed:b:oed},
which costs \$3000 in print form, but is a (relative) bargain on
CD-ROM.

\subsection{Language Usage Handbooks/Guides}\label{s:guides}

A quick visit to any bookstore will reveal that this is a hot 
area with many choices.  One that is our favorite, and that is
sure to warm the (algorithmic) heart of any computer scientist,
is Johnson's marvelous book {\em The Handbook of Good 
English}~\cite{joh:b:handbook}.  If you want to know why
this sentence:
\begin{quote}
The polynomial-time machines run in polynomial time.
\end{quote}
is correct (note the hyphens), or what the difference is between ``the
friends of John'' and ``the friends of John's,'' Johnson's book is a
wonderful place to find clear and fascinating answers.  His book will
even tell you why we put the comma inside the quotation marks in the previous
sentence.

Another favorite is 
van~Leunen's 
{\em A Handbook for Scholars}~\cite{leu:b:scholars}.  
This well-written book is
highly recommended by 
Knuth, Larrabee, and Roberts~\cite{knu-lar-rob:b:math-writing}, whose guide on
mathematical writing we will recommend  below.  Although this is a general
purpose handbook for scholars of all kinds, van~Leunen seems to have special 
insight into the ordinary writing errors that seem  
most prevalent among computer scientists. In addition, we know of no other
guide that name-drops our colleagues:  Look for mention of 
Zalcstein, Ullman, and Aho.  Read about Ramanujan and the great 
mathematician G.\ H. Hardy.  We recommend this source especially for its
discussion of consistent bibliography style and meaningful citations.

The famous book by Strunk and White, 
{\em The Elements of Style}~\cite{str-whi:b:elements},
also falls into this category.  
{\em The Elements of Style\/} is amazingly
short, and can be read in a flash.  Most of us probably still have
the copy we were required to purchase as students.

{\em Merriam-Webster's Dictionary of English
Usage}~\cite{mer:b:english-usage} (yes, that really is its title)
provides a sharp contrast to Strunk and White.  
{\em Merriam-Webster's Dictionary of English
Usage\/} is huge, and if
anything is too willing to embrace usage shifts.

Speaking of huge books, {\em The 
Chicago Manual of Style}~\cite{chi:b:manual-style}
provides a guide through every aspect of the writing 
and editing process.  There is so much information here that
it can be a bit hard to find what one is looking for, so 
we usually reach first for Johnson or van Leunen.  However, 
{\em The Chicago Manual of Style\/} is 
certainly a useful reference to have on one's shelf.

These books differ on the advice they give on some issues.  
For example, consider the difference between ``that'' and ``which.''
Many of us use these words correctly when we are talking, but 
indiscriminately substitute ``which'' for ``that'' when we are writing.
There is an easy rule---the word
``which'' should 
come after only commas or prepositions when it is being used 
to introduce a nonrestrictive relative 
clause---that seems to work in 
most situations.   Even so,  there are many exceptions, for which,
as always, one should refer to the excellent guides that are available.
Van Leunen contains a section called 
```Which'-Hunting,'' and Strunk and White is inflexible on the issue.
On the other hand, {\em Merriam-Webster's Dictionary of English
Usage\/} provides a very rich historical discussion of the issue, 
followed by reasonable
(though quite liberal) advice.  On almost each issue, it 
both lets one know where the experts fall, and then gives specific
advice as to what rule one should follow.
It is true enough,
as {\em Merriam-Webster's Dictionary of English
Usage\/} mentions,
that Shakespeare's plays do not 
follow the rule favored by Strunk and White.
Technical papers, however, usually do not convey the depths of 
human passion.  We are creating  neither comedy nor tragedy.  Our goal
is to accurately describe complicated technical material without
confusing or losing our readers along the way.  For this reason, we
suggest following those rules and conventions that 
best avoid ambiguity.  To return to our example,
the Strunk and White approach to ``that'' and ``which'' 
allows writers to clearly express distinct
meanings to their readers.  Similarly, adopting the 
Final Serial Comma Convention---using a comma before the ``and'' 
in lists of length at least three---helps avoid 
ambiguity (see also Section~\ref{ss:wtcs}).

\subsection{Chatty Books on Language Usage}

The books of Section~\ref{s:guides} are not exactly ones that  lend
themselves to being read.  Rather, they are excellent reference books.
However, there are a large number of books about language that one can
actually enjoy reading.  
Safire's books on 
language are true joys ({\em Coming to Terms};  {\em Fumblerules};
{\em Language Maven Strikes Again};  {\em You Could Look It Up};
{\em Take My Word for It}; {\em I Stand Corrected: More on 
Language}; {\em What's the Good Word?}; and
{\em On Language}), as are those of 
Theodore
Bernstein.

\subsection{Reverse Spelling Dictionaries}

These list words under their common misspellings.  This is a cute
idea, but that's about all it is.  Ignore them and instead use a
spelling checker that suggests corrections (ispell, Excalibur, etc.).

\subsection{Writing Elegantly}

Would that one could read a book and come away an elegant writer.
However, we enthusiastically commend to the reader the charming book
{\em Style:  Ten Lessons in Clarity and Grace}~\cite{wil:b:style}.  
At 
a more introductory level, we have found 
{\em The Lively Art of Writing\/} to be very helpful.

\subsection{Writing (Theoretical) Computer Science}\label{ss:wtcs}
There is a small but growing collection of books focusing on
how computer scientists should write.
There is a huge 
literature on technical writing.  Our own view is that 
the best way to be an excellent writer of computer science is to 
be an excellent writer.    Nonetheless, 
there are  issues of specific interest to
theoretical
computer scientists.  
Among these issues are how to format equations, and how to
unambiguously express quantifiers in English text (good
luck).  

One of the best books addressing such issues is by
Knuth, Larrabee, and Roberts~\cite{knu-lar-rob:b:math-writing}.
Their book includes a clear
list of rules.   Some of these
are obvious, such as ``Don't start a sentence with a symbol,'' and others
discuss more subtle aspects of mathematical style.  

Here we mention one subtle point about typesetting computer science that is
often missed.  In English, it is fine to ignore the Final Serial Comma
Convention (e.g., to write ``A, B and C'') throughout a paper.  It is
also fine to use it (e.g., to write ``A, B, and C'') throughout a
paper.  However, one should be consistent.  This is a bit harder than
one might expect.  The reason is that Bib\TeX~styles have to make a
choice about this.  Most of the common ones (wisely) choose to use the
convention.  So, when using Bib\TeX,~one should make sure to adopt in
the body of the paper the same convention Bib\TeX~is using in the
bibliography of the paper.

The very best book we know on writing theoretical computer science 
isn't even about writing theoretical computer science.  The book---whose 
stylized title
does not adopt the Final Serial Comma Convention---is 
Krantz's {\em 
A Primer of Mathematical Writing: Being a Disquisition on
Having Your Ideas Recorded, Typeset, Published, Read \&
Appreciated}~\cite{kra:b:writing}, and it is amazing.  Going far beyond the
small technical details of writing, Krantz discusses how to write grants,
reference letters, and so on.  His section on how to write and read
tenure-case letters is a show-stopper.

\subsection{Specialized Resources and Fun Toys}

We have on our shelves some fun toys.  
{\em The {BBI} Combinatory 
Dictionary of English}~\cite{ben-ben-ils:b:combinatory}
does just what its title
promises.  It tells you what mortar to use to glue together different
world combinations.  
Ehrlich's 
{\em Amo, Amas, Amat and More: How to Use Latin to Your Own
Advantage and to the Astonishment of Others}~\cite{ehr:b:amas-latin}
helps one decode (or use) the 
bits of Latin that often wander into English usage.  Finally, just 
for sheer perversity, Hunsberger's
{\em The Quintessential Dictionary}~\cite{hun:b:quintessential}
gives you a stockpile of words sure to 
delight you and distress your readers.  You'd have to crazy to put
any of them into a technical paper, but perhaps you'll sleep better
(though less) if you know what ``syndyasmian'' means.  Off course,
one of the standard quotation books (for example, 
{\em Bartlett's Familiar 
Quotations}~\cite{bar:b:quotations}) 
can be useful when hunting for the right quotation.
\begin{quote}
Nothing is given so profusely as advice.---Fran\c{c}ois, Duc
de La Rochefoucauld, {\em Reflections; or, Sentences and Moral
Maxims}, 1678.
\end{quote}

\section{Common Problems}
The following list contains typical items that we looked for while editing
and refers to errors that occurred frequently often to deserve 
attention.  We will keep these items brief, and allow our readers 
to study the various sources of Section~\ref{s:tools} 
for complete descriptions:  

\begin{description}
\item[Correct use of ``that'' and ``which'']\quad
\item[Correct use of hyphens] \quad  For example, the following sentence is
hyphenated correctly:  
\begin{quote} A polynomial-time machine runs in polynomial time.
\end{quote}

\item [Correct punctuation of quotations] \quad
As our book used American spelling, we also used the 
American convention for the location of quotation marks.  
For example, in American English this:
\begin{quote}
If Mary says ``ternary,'' then yell ``binary.''
\end{quote}
is correct, as is the following (really---``!''~and ``.''~fall
into different equivalence classes in this regard):
\begin{quote}
If Mary says ``ternary,'' then yell ``binary''!
\end{quote}
However in British English, these would be written, quite logically, as:
\begin{quote}
If Mary says ``ternary'', then yell ``binary''.
\end{quote}
and
\begin{quote}
If Mary says ``ternary'', then yell ``binary''!
\end{quote}
The American convention is due to action of the metal slugs 
used in the type of printing presses one sees these days only at 
museums and historical recreations.  However, there is something 
quite sweet in seeing laser printers printing pages whose conventions
are shaped by a much earlier technology.

\item[Complete bibliographies]\quad
Bibliographies should be complete.  For example, journal names
and the names of conference proceedings should be spelled out.
\item[Citation style]\quad
The modern style of scholarly citations, as every \LaTeX~user knows, 
uses pointers, enclosed in brackets, that refer to items in a list of 
references.
This style replaced an older more cumbersome practice that used extensive 
footnotes.  However, the pointers themselves should not be part of
an article's text.  That is, ``Baker, Gill, and Solovay [BGH75] noticed that''
is better than ``[BGS75] noticed that.''
\item[No hardcoded cross-references]\quad 
As we stated above, cross-references should not be hardcoded.
\item[Correct punctuation of equations]\quad
Equations should be correctly punctuated.  For example, if a displayed 
equation ends a sentence, then it should end with a period.
\item[Consistent theorem format]\quad
Theorem format should be consistent.  For example,  definitions should
not appear sometimes in italic and other times in roman.
\end{description}

\section{Conclusions}

In this article, we have recounted some of our experiences as editors,
and we have listed some books that we have found to be very helpful to
us when writing and editing.  We emphasize that this article presented
our personal preferences, based on the very finite set of books we
have been exposed to.  Your exposure, preferences, and mileage may
differ.

\bibliographystyle{alpha}


\begin{thebibliography}{OED89}

\bibitem[AH91]{ah:b:dictionary}
{\em The {A}merican {H}eritage Dictionary of the {E}nglish Language}.
\newblock Houghton Mifflin, 3rd edition, 1991.

\bibitem[Bar92]{bar:b:quotations}
J.~Bartlett.
\newblock {\em Bartlett's Familiar Quotations: {A} Collection of Passages,
  Phrases, and Proverbs Traced to Their Sources in Ancient and Modern
  Literature}.
\newblock Little, Brown, and Company, 16th edition, 1992.

\bibitem[BBI86]{ben-ben-ils:b:combinatory}
M.~Benson, E.~Benson, and R.~Ilson.
\newblock {\em The {BBI} Combinatory Dictionary of English: {A} Guide to Word
  Combinations}.
\newblock Benjamins, 1986.

\bibitem[Chi93]{chi:b:manual-style}
{\em The {C}hicago Manual of Style: The Essential Guide for Writers, Editors,
  and Publishers}.
\newblock University of Chicago Press, 14th edition, 1993.

\bibitem[Ehr93]{ehr:b:amas-latin}
E.~Ehrlich.
\newblock {\em Amo, Amas, Amat and More: {H}ow to Use {L}atin to Your Own
  Advantage and to the Astonishment of Others}.
\newblock HarperCollins, 1993.

\bibitem[HS97]{hem-sel:b:ctrII}
L.~Hemaspaandra and A.~Selman, editors.
\newblock {\em Complexity Theory Retrospective II}.
\newblock Springer-Verlag, 1997.

\bibitem[Hun84]{hun:b:quintessential}
I.~Hunsberger.
\newblock {\em The Qunitessential Dictionary}.
\newblock Warner Books, 1984.

\bibitem[Joh91]{joh:b:handbook}
E.~Johnson.
\newblock {\em The Handbook of Good English: {A} Comprehensive, Easy-to-Use
  Guide to Modern Grammar, Punctuation, Usage, and Style}.
\newblock Washington Square Press, revised and updated edition, 1991.

\bibitem[KLR89]{knu-lar-rob:b:math-writing}
D.~Knuth, T.~Larrabee, and P.~Roberts.
\newblock {\em Mathematical Writing}.
\newblock Mathematical Association of America Notes Number 14. Mathematical
  Association of America, 1989.

\bibitem[Kra97]{kra:b:writing}
S.~Krantz.
\newblock {\em A Primer of Mathematical Writing: {B}eing a Disquisition on
  Having Your Ideas Recorded, Typeset, Published, Read \& Appreciated}.
\newblock AMS Press, 1997.

\bibitem[MW34]{web:b:new-international-second-edition}
{\em Webster's New International Dictionary (Unabridged)}.
\newblock G \& C. Merriam Company, 2nd edition, 1934.

\bibitem[MW93]{web:b:new-international-third-edition}
{\em Webster's Third New International Dictionary (Unabridged)}.
\newblock Merriam-Webster, 1993.
\newblock P.~Gove, editor.

\bibitem[MW94]{mer:b:english-usage}
{\em Merriam-Webster's Dictionary of English Usage}.
\newblock Merriam-Webster, Inc., 1994.

\bibitem[OED89]{oed:b:oed}
{\em Oxford English Dictionary}.
\newblock Oxford University Press, 2nd edition, 1989.
\newblock P.~Gove, editor.

\bibitem[RH93]{fle:b:random-dictionary}
{\em {R}andom {H}ouse Unabridged Dictionary}.
\newblock Random House, 1993.
\newblock 2nd edition, newly revised and updated, S. Flexner, editor.

\bibitem[Sel90]{sel:b:retrospective}
A.~Selman, editor.
\newblock {\em Complexity Theory Retrospective}.
\newblock Springer-Verlag, 1990.

\bibitem[SW79]{str-whi:b:elements}
W.~{Strunk, Jr.} and E.~White.
\newblock {\em The Elements of Style}.
\newblock Macmillan Publishing Company, New York, 3rd edition, 1979.

\bibitem[vL92]{leu:b:scholars}
Mary-Claire van Leunen.
\newblock {\em A Handbook for Scholars}.
\newblock Oxford University Press, revised edition, 1992.

\bibitem[Wil96]{wil:b:style}
J.~Williams.
\newblock {\em Style: Ten Lessons in Clarity and Grace}.
\newblock Addison-Wesley, 5th edition, 1996.

\end{thebibliography}

\end{document}